\documentclass[prd,amsmath,amssymb,superscriptaddress,nofootinbib]{revtex4}

\usepackage{graphicx}
\usepackage{dcolumn}
\usepackage{bm}

\normalbaselines \topmargin -0.25 truein \oddsidemargin 0.30
truein \evensidemargin 0.30 truein 

\newcommand{\D}{\hat{D}}

\begin{document}

\title{SU(N) - symmetric dynamic aether: \\ General formalism and a hypothesis on spontaneous color polarization}

\author{Alexander B. Balakin}
\email{Alexander.Balakin@ksu.ru} \affiliation{Department of General
Relativity and Gravitation, Institute of Physics, Kazan Federal University, Kremlevskaya
str. 18, Kazan 420008, Russia}
\author{Alexey V. Andreyanov}
\email{andreyanov.alexey.27@gmail.com} \affiliation{Department of General
Relativity and Gravitation, Institute of Physics, Kazan Federal University, Kremlevskaya
str. 18, Kazan 420008, Russia}

\date{\today}

\begin{abstract}

The SU(N)-symmetric generalization of the model of the electromagnetically active dynamic aether is formulated.
This generalization is based on the introduction of a Yang-Mills gauge field instead of the Maxwell field, and of a SU(N)-multiplet of vector fields instead of the standard single vector field.
In the framework of the second order version of the effective field theory this generalization includes three constitutive tensors, which are the SU(N) extensions of the tensors appeared in the Einstein-Maxwell-aether theory; we reconstructed the full-format set of these constitutive tensors.
The total self-consistent system of master equations for the gauge, vector and gravitational fields is obtained by the variation procedure. The general model of the SU(N)-symmetric dynamic aether is reduced to the extended Einstein-Yang-Mills-aether model by the ansatz about spontaneous color polarization of the vector fields. In fact, this ansatz requires the vector fields, which  form the SU(N) multiplet, to become parallel in the group (color) space due to a phase transition, and a new selected direction in the group space to appear, thus converting it into the anisotropic color space.

\end{abstract}

\maketitle

\section{Introduction}\label{Intro}

The term {\it dynamic aether} is associated with the modification of the theory of gravity, which is based on the introduction of a  time-like unit dynamic vector field $U^i$ interpreted as the velocity four-vector of some global substratum, the aether. In this sense, the Einstein-aether theory  \cite{J1,J2,J3,J4,J5,J6,J7,J8,J9,J10} belongs to the class of vector-tensor theories of gravity, which forms the corresponding branch in the modern science entitled as Modified Theories of Gravity (see, e.g., \cite{O1,O2}). The vector field $U^i$ is unit and time-like,  $g_{ik}U^i U^k =1 >0$, since  the Einstein-aether theory is considered as a realization of the idea of a preferred frame of reference \cite{CW,N1,N2}, associated with a world-line congruence, for which the corresponding time-like velocity four-vector is the tangent vector. The vector field $U^i$ is indicated as the dynamic one, since the Lagrangian of the theory contains quadratic combinations of the covariant derivative $\nabla_k U_i$, and, correspondingly, the master equations for the vector field are of the second order in derivatives \cite{J1}. There are three extensions of the Einstein-aether theory. The first extension deals with a supplementary pseudoscalar field and is indicated as Einstein-aether-axion model  \cite{B2016}. The second extension includes the electromagnetic field and is entitled as  Einstein-Maxwell-aether theory \cite{BL2014}. The third extension is the Einstein-Maxwell-aether-axion theory \cite{AB2016}. In these extended theories the dynamic aether is considered as a quasi-medium, and the unit vector field $U^i$ plays the role of the global velocity of this quasi-medium. Since, generally, the velocity field is non-uniform, various effects induced in the dynamic aether are predicted based on the analogy with electrodynamic phenomena in classical moving media (see, e.g., the review \cite{Symm}, and \cite{Ap1,Ap2,Ap3,Ap4}). The Einstein-Maxwell-aether theory uses two vector fields: the first one is the unit dynamic vector field $U^i$; the second is presented by the electromagnetic potential four-(co)vector, $A_i$.
Only the skew-symmetric part of the derivative of the potential, $F_{ik}=\nabla_iA_k{-}\nabla_kA_i$, the Maxwell tensor, enters the Lagrangian of the  Einstein-Maxwell-aether theory \cite{BL2014};
the four-vector $A_i$ itself does not appear in the Lagrangian thus providing the $U(1)$-gauge invariance of the theory.

The next step, which we do, is the SU(N) - symmetric generalization of the dynamic aether theory. This step is motivated by the fact that the relativistic cosmology and astrophysics offer a lot of interesting problems appropriated just for the theory of non-Abelian gauge fields (see, e.g., the review \cite{VG}). Clearly, the non-Abelian version of the dynamic aether theory also could attract the attention of physicists. We consider now the full-format SU(N)-symmetric dynamic aether, i.e., we introduce the Yang-Mills gauge field instead of the U(1)-symmetric Maxwell field, and the multiplet of vector fields $\{U^{i(a)}\}$ instead of the single vector field $U^i$. Then, in order to return to the paradigm of global velocity field of the aether, we use the idea of spontaneous color polarization mechanism, which reduces the color multiplet $\{U^{i(a)}\}$ to the product $U^{i(a)}= q^{(a)}U^i$, which contains the velocity four-vector $U^i$ and the multiplet of scalars $q^{(a)}$ forming the vector in the group space (for the sake of simplicity, below we use the term {\it color vector} for this vector in the group space). Clearly, this is the idea of formation of the set of vector fields {\it parallel} in the group space. Such idea is similar to the one used for the gauge potentials (see, e.g., \cite{Yasskin,G,Parametric}). Parallel vector fields $U^{i(a)}= q^{(a)}U^i$ could appear as a result of some phase transition. Let us emphasize, that mentioned spontaneous color polarization involves the multiplet of vector fields; the gauge field can remain the non-Abelian one, or can be also exposed by parallelization. In the last situation the gauge potentials $A^{(b)}_m$ convert into the quasi-Abelian set of parallel potentials $A^{(b)}_m = Q^{(b)}A_m$. The color vectors $q^{(a)}$ and $Q^{(a)}$ can coincide or can be different; however, in any cases the group space becomes anisotropic. Based on the analogy with electrodynamics of continuous media \cite{LL,BZ}, one can indicate the model with only one color vector ($q^{(a)}$, or $Q^{(a)}$, or $q^{(a)}=Q^{(a)}$) as the model with uni-axial color space; respectively, the model with $q^{(a)}\neq Q^{(a)}$ can be indicated as the bi-axial one. In the framework of the non-minimal Einstein-Yang-Mills-Higgs theory the anisotropic color spaces were considered in \cite{a1,a2,a3,a4}; in that models the anisotropy of the group space was assumed to be induced by the multiplet of Higgs scalar fields $\Phi^{(a)}$, and now such anisotropy is connected with the multiplet of vector fields $U^{i(a)}$.

The purpose of this paper is to formulate the mathematical foundations of the new theory, which describes the SU(N)-symmetric dynamic aether. The paper is organized as follows.
In Section II we recall the basic formalism of the Einstein-aether and of the Einstein-Maxwell-aether theories. Section III contains the basic elements of the formalism of the SU(N)-symmetric model. In Section IV we discuss the simplest version of the SU(N)-symmetric dynamic aether theory, namely, the Einstein-Yang-Mills-aether model, in which the vector field $U^i$ is single, and the group space is isotropic. In Section V we establish the full-format theory of the SU(N)-symmetric dynamic aether, and reconstruct three basic color constitutive tensors, which appear in the framework of the second order version of the effective field theory \cite{EFT1,EFT2}. In Section VI we discuss the formalism of spontaneous color polarization and its consequence for the structure of the reduced master equations. Conclusions are presented in Section VII.

\section{Preamble}\label{sec1}

\subsection{Formalism of the Einstein-aether theory}

The Einstein-aether theory is based on the action functional
\begin{equation}\label{01act}
S_{(\rm EA)} = \int d^4 x \sqrt{-g}\ \left\{ \frac{1}{2\kappa}\left[R + 2\Lambda + \lambda\left(g_{mn}U^{m}U^{n} -1 \right) + K^{ijmn} \nabla_i U_{m} \nabla_j
U_n \right] \right\} \,,
\end{equation}
(see, e.g., \cite{J1}). Here three standard elements of the Einstein-Hilbert action are introduced, namely, the determinant of the metric $g {=} {\rm det}(g_{ik})$, the Ricci
scalar $R$, the cosmological constant $\Lambda$.
There are also two new terms involving the vector field $U^i$. The
first term $\lambda \left(g_{mn}U^m U^n {-}1 \right) $ guarantees that the $U^i$ is normalized to one. The second term $K^{ijmn} \
\nabla_i U_m \ \nabla_j U_n $ is quadratic in the covariant derivative
$\nabla_i U_m $ of the vector field; the constitutive tensor $K^{ijmn}$ is constructed
using the metric tensor $g^{ij}$ and the velocity four-vector $U^k$ as follows:
\begin{equation}
K^{ijmn} {=} C_1 g^{ij} g^{mn} {+} C_2 g^{im}g^{jn}
{+} C_3 g^{in}g^{jm} {+} C_4 U^{i} U^{j}g^{mn}\,.
\label{2}
\end{equation}
Four parameters $C_1$, $C_2$, $C_3$ and $C_4$ are the Jacobson constants \cite{J1,J2,J3}.

The action functional (\ref{01act}) contains three quantities attributed for variation procedure: the Lagrange multiplier $\lambda$, the vector field $U^i$ and the space-time metric $g^{ij}$.
The variation with respect to $\lambda$ yields the normalization condition for the time-like vector field $U^k$.
\begin{equation}
g_{mn}U^m U^n = 1 \,.
\label{21}
\end{equation}
The result of variation of the functional (\ref{01act}) with respect to
$U^i$ can be written in the form
\begin{equation}
\nabla_m {\cal J}^{mn}_{({\rm A})}
- I^n_{({\rm A})} = \lambda \ U^n  \,,
\label{0A1}
\end{equation}
where
\begin{equation}
{\cal J}^{mn}_{({\rm A})} =  K^{lmsn} \nabla_l U_s \,,
\label{0A31}
\end{equation}
\begin{equation}
I^n_{({\rm A})} =  \frac12 \nabla_l U_s \nabla_m U_j \
 \frac{\delta K^{lsmj}}{\delta U_n}  =  C_4  U^p \nabla_p U_m \nabla^n U^m
\,.
\label{0A4A}
\end{equation}
Convolution of (\ref{0A1}) with the velocity four-vector gives the Lagrange multiplier
\begin{equation}
\lambda = U_n \left[\nabla_m {\cal J}^{mn}_{({\rm A})}
- I^n_{({\rm A})}\right]  \,.  \label{0A309}
\end{equation}
Excluding $\lambda$ from (\ref{0A1}) we obtain
\begin{equation}
\Delta_n^j \left[\nabla_m {\cal J}^{mn}_{({\rm A})} -
I^n_{({\rm A})} \right] = 0 \,,
\label{0A19}
\end{equation}
where $\Delta_n^j$ is the projector
$\Delta_n^j \equiv \delta_n^j {-}U_n U^j$.

The variation of the action (\ref{01act}) with respect to the metric
gives the gravitational field equations in the form
\begin{equation}
R_{ik} - \frac{1}{2} R \ g_{ik} -  \Lambda g_{ik} = T^{({\rm U})}_{ik}
\,.
\label{0Ein1}
\end{equation}
The term $T^{({\rm U})}_{ik}$ presents the stress-energy tensor of the
of the vector field $U^i$:
$$
T^{({\rm U})}_{ik} = C_1\left(\nabla_mU_i \nabla^m U_k {-}
\nabla_i U_m \nabla_k U^m \right) {+} C_4 U^p \nabla_p U_i U^q \nabla_q U_k {+}
$$
\begin{equation}
{+}\frac12 g_{ik} {\cal J}^{am}_{({\rm A})} \nabla_a U_m {+}
\nabla^m \left[U_{(i}{\cal J}_{k)m}^{({\rm A})}\right] {-}
\nabla^m \left[{\cal J}_{m(i}^{({\rm A})}U_{k)} \right] {-}
\nabla_m \left[{\cal J}_{(ik)}^{({\rm A})} U^m\right] {+} U_iU_k
U_n \left[\nabla_m {\cal J}^{mn}_{({\rm A})} {-} I^n_{({\rm A})}
\right]\,, \label{0Ein5}
\end{equation}
where $p_{(i} q_{k)}{\equiv}\frac12 (p_iq_k{+}p_kq_i)$
denotes
symmetrization.  The tensor $T^{({\rm U})}_{ik}$ disappears when
the motion of the aether is uniform, i.e., $\nabla_iU_k{=}0$.

\vspace{3mm}
\noindent
{\it Remark}

\noindent
As usual, here we omit the terms of the type $\Psi^k U^n \nabla_k U_n$, which could be introduced, formally speaking, into the action functional (\ref{01act}) ($\Psi^m$ is arbitrary function of $U^j$ and $\nabla_jU^s$).
In the framework of Einstein-aether and Einstein-Maxwell-aether theories, such terms after integration by parts can be reduced to $-\frac12 (g_{mn}U^m U^n-1)\nabla_k \Psi^k$. Clearly, this leads to the redefinition of the Lagrange multiplier $\lambda \to \lambda^{*} {=} \lambda {-}\frac12 \nabla_k \Psi^k$, only, i.e., the master equations keep the form. Below we will show that in the theories with SU(N) symmetry the terms of the mentioned type can not be neglected.

\subsection{Formalism of the Einstein-Maxwell-aether theory}

\subsubsection{The Lagrangian}

The U(1)-symmetric extension of the Einstein-aether theory is associated with introduction of the gauge invariant Maxwell tensor $F_{ik}$ into the action functional. The Maxwell tensor is  the skew-symmetrized derivative of the potential four-vector
\begin{equation}
F_{ik}{=}\nabla_i A_k {-}\nabla_k A_i\,.
\label{maxwellA1}
\end{equation}
We consider now the terms of the second order in the derivatives according to the principles of effective field theories (see,
e.g., \cite{EFT1,EFT2}); this means that we can use the second-order terms of three types: quadratic in the covariant derivative of the vector field, quadratic in the Maxwell tensor, and cross-terms containing the product of mentioned tensors. The corresponding new contributions into the action functional (additional to (\ref{01act}))  are
\begin{equation}\label{02act}
S_{(\rm EM)} = \int d^4 x \sqrt{-g}\ \left\{ \frac{1}{2} A^{mnik}F_{ik} \nabla_m U_{n} + \frac{1}{4} C^{ikmn}F_{ik} F_{mn}  \right\}\,.
\end{equation}
Here the coefficients $A^{ikpq}$ describe effects of spontaneous polarization-magnetization induced by the dynamic aether. The representation of the tensor $A^{mnik}$ is discussed in \cite{BL2014};
it can be reconstructed using the metric $g_{ik}$, the covariant constant Kronecker tensors
($\delta^i_k$, $\delta^{ik}_{ab}$ and higher order Kronecker tensors),
the Levi-Civita tensor $\epsilon^{ikab}$, the unit vector field
$U^k$, and a pair of coupling constants. The coupling constants $\pi_1$ and $\mu_1$ were interpreted in \cite{BL2014} as coefficients describing the polarization and magnetization of the aether, which are induced by the aether non-uniform motion.
Keeping in mind further SU(N) generalization of the Einstein-Maxwell-aether theory, we consider here the following motives for the Lagrangian decomposition.

\vspace{3mm}
\noindent
{\it First}, there are three irreducible invariants linear in the Maxwell tensor $F_{ik}$ and linear in the tensor $\nabla_m U_n$, which are constructed using metric $g^{js}$ and $U^l$
\begin{equation}
g^{im} g^{kn}F_{ik} \nabla_m U_n \,, \quad g^{in} U^k U^m F_{ik} \nabla_m U_n  \,, \quad g^{im} U^k U^n F_{ik} \nabla_m U_n   \,.
\label{EMA1}
\end{equation}
The last invariant (see Remark) can be reduced to $-\frac12 (g_{mn}U^mU^n-1)\nabla_i (F^{ik}U_k)$ using the integration by parts. In other words, we deal again with the redefinition of the Lagrange multiplier.

\vspace{3mm}
\noindent
{\it Second}, we can introduce two coupling parameters $\omega_1$ and $\omega_2$ and can write
\begin{equation}
\frac{1}{2} A^{mnik}F_{ik} \nabla_m U_{n} = \frac12 F_{ik} \left(\omega_1 g^{im} g^{kn}  + \omega_2 g^{in} U^k U^m \right)\nabla_m U_n    \,.
\label{EMA2}
\end{equation}
Similarly, there are only two irreducible terms containing the Maxwell tensor in square in composition with metric and the velocity four-vector:
\begin{equation}
g^{im} g^{kn}F_{ik} F_{mn}  \,, \quad g^{im} U^k U^n F_{ik} F_{mn} \,.
\label{EMA3}
\end{equation}
The corresponding term in the Lagrangian is of the form
\begin{equation}
\frac{1}{4} C^{ikmn}F_{ik} F_{mn}  =  \frac{1}{4 \mu} \left[F_{ik}F^{ik} {+} 2(\varepsilon \mu
{-}1)F_{im}U^m F^{in}U_n \right]  \,.
\label{M20}
\end{equation}
According to the standard interpretation, $\varepsilon$ is the the dielectric permittivity of the aether, and $\mu$ is the magnetic permeability. The linear response tensor $C^{ikmn}$ can be written in the form
\begin{equation}
C^{ikmn} = \frac{1}{2\mu}\left[
g^{im} g^{kn}{-}g^{in} g^{km} + (\varepsilon \mu {-}1)\left(g^{im} U^kU^n {-} g^{in} U^kU^m {+}g^{kn} U^iU^m {-}g^{km} U^iU^n \right)\right] \,,
\label{M2b}
\end{equation}
or equivalently,
\begin{equation}
2C^{ikmn} =  \varepsilon g^{ikmn} + \left(\frac{1}{\mu}-\varepsilon \right) \Delta^{ikmn}  \,,
\label{M209}
\end{equation}
where the auxiliary tensors are defined as follows:
\begin{equation}
g^{mnpq} \equiv g^{mp} g^{nq}{-}g^{mq} g^{np} \,, \quad
\Delta^{mnpq} \equiv \Delta^{mp} \Delta^{nq}{-}\Delta^{mq}
\Delta^{np} \,.
\label{M321}
\end{equation}
In these terms the tensor $A^{mnik}$ has the form
\begin{equation}
A^{mnik} = \frac12 (\omega_2-\omega_1) g^{iknl} U^m U_l + \frac12 \omega_1 \Delta^{ikmn} \,,
\label{M5}
\end{equation}
i.e., the constants $\pi_1$ and $\mu_1$ introduced in \cite{BL2014} are connected with $\omega_1$ and $\omega_2$ as follows: $\pi_1 = \frac12(\omega_2-\omega_1)$, $\mu_1 = - \frac12 \omega_1$.

The total action functional of the Einstein-Maxwell-aether theory in the second order of the effective field theory is the sum
$S_{(\rm EMA)} = S_{(\rm EA)} + S_{(\rm EM)}$:
$$
S_{(\rm EA)} = \int d^4 x \sqrt{-g}\ \left\{ \frac{1}{2\kappa}\left[R + 2\Lambda + \lambda\left(g_{mn}U^{m}U^{n} -1 \right) + K^{ijmn} \nabla_i U_{m} \nabla_j
U_n \right] + \right.
$$
\begin{equation}\label{totalact}
\left. + \frac{1}{2} A^{mnik}F_{ik} \nabla_m U_{n} + \frac{1}{4} C^{ikmn}F_{ik} F_{mn}\right\} \,,
\end{equation}
and now we are ready to start the variation procedure.

\subsubsection{Electrodynamic equations}
\label{electroeq}

The electrodynamic equations is the result variation of the total action functional with
respect to the electromagnetic potential four-vector $A_i$. This result can be written
as follows:
\begin{equation}
\nabla_k H^{ik} = 0 \,, \quad H^{ik} = {\cal H}^{ik} + C^{ikmn}F_{mn} \,,
\label{E2}
\end{equation}
where $H^{ik}$ is the excitation tensor linear in the Maxwell tensor, and the tensor
\begin{equation}
{\cal H}^{ik} =  A^{mnik} \nabla_m U_n  \label{E4}
\end{equation}
describes the spontaneous polarization-magnetization of
the non-uniformly moving aether.
Also, we have to add the relationships
\begin{equation}
\nabla_k F^{*ik} = 0 \,,
\label{E9}
\end{equation}
where the asterisk indicates the standard dualization procedure
\begin{equation}
F^{*ik} = \frac{1}{2} \epsilon^{ikmn} F_{mn} \,.
\label{E10}
\end{equation}
Here $\epsilon^{ikmn} {=} \frac{{\rm E}^{ikmn}}{\sqrt{{-}g}}$ is the
Levi-Civita tensor, ${\rm E}^{ikmn}$ is the completely skew-symmetric symbol with ${\rm E}^{0123}{=}1$.

\subsubsection{Dynamic equations of the electromagnetically active aether}
\label{aethereqs}

The variation of the total action functional with respect to the vector field yields
\begin{equation}
\nabla_m \left[{\cal J}^{mn}_{({\rm A})} {+} \kappa {\cal
J}^{mn}_{({\rm M})}\right] = I^n_{({\rm A})} {+} \kappa I^n_{({\rm M})} {+} \lambda \ U^n \,.
\label{5A1}
\end{equation}
Here ${\cal J}^{mn}_{({\rm M})}$ and $I^n_{({\rm M})}$ are given, respectively, by
\begin{equation}
{\cal J}^{mn}_{({\rm M})} =  \frac{1}{2} \left[\omega_1 F^{mn} + \omega_2 F^{nk}U_k U^m \right] \,,
\label{5A61}
\end{equation}
$$
I^n_{({\rm M})} \equiv \frac12 \left[\frac{\delta A^{pqik}}{\delta U_n} \right] F_{ik} \nabla_p U_q + \frac14 \left[\frac{\delta C^{ikpq}}{\delta U_n} \right] F_{ik}F_{pq} =
$$
\begin{equation}
 =  \frac12 \omega_2 \left( F^{pq}U_q  \nabla^n U_p + F^{pn} U^l \nabla_l U_p \right) + \left(\varepsilon {-}\frac{1}{\mu} \right)
F^{kn}F_{km} U^m
\,. \label{5A62}
\end{equation}
The Lagrange multiplier is equal to the sum
\begin{equation}
\lambda = U_n \left[\nabla_m {\cal J}^{mn}_{({\rm A})} -
I^n_{({\rm A})} \right] + \kappa U_n \left[\nabla_m {\cal J}^{mn}_{({\rm M})} -
I^n_{({\rm M})} \right] \,,
\label{5A30-1}
\end{equation}
and we finally obtain
\begin{equation}
\Delta_n^s \left\{\nabla_m \left[{\cal J}^{mn}_{({\rm A})} {+} \kappa
{\cal J}^{mn}_{({\rm M})}\right] {-} \left[I^n_{({\rm A})} {+} \kappa I^n_{({\rm M})} \right] \right\} = 0 \,.
\label{A2}
\end{equation}

\subsubsection{Master equations for the gravitational field}
\label{graveqs}

The variation of the total action functional with respect to
the metric $g^{ik}$ yields
\begin{equation}
 R_{ik} {-} \frac{1}{2} R \ g_{ik} {-}  \Lambda g_{ik} =
 T^{({\rm U})}_{ik} + \kappa \left[T^{({\rm EM})}_{ik}{+} T^{({\rm
EMA})}_{ik} \right] \,.
\label{Ein1}
\end{equation}
The term $T^{({\rm U})}_{ik}$ is already presented above by the formula
(\ref{0Ein5}).
The term indicated as $T^{({\rm EM})}_{ik}$ is given by
$$
T^{({\rm EM})}_{ik} {=} \frac{1}{\mu} \left\{\left[\frac14
g_{ik}F_{mn}F^{mn}{-}F_{im}F_k^{\ m}\right] {+} \right.
$$
\begin{equation}
\left. +(\varepsilon \mu {-}
1)U^p U^q\left[\left(\frac12 g_{ik}{-} U_iU_k\right) F^{m}_{\ \ p}
F_{mq} {-} F_{ip}F_{kq} \right] \right\}
\,. \label{5Ein6}
\end{equation}
In the vacuum, when $\varepsilon {=} \mu {=} 1$, the tensor $T^{({\rm EM})}_{ik}$
gives the usual Maxwell term.
The tensor (\ref{5Ein6}) is symmetric and traceless,
i.e.,
\begin{equation}
T^{({\rm EM})}_{ik} =  T^{({\rm EM})}_{ki} \,, \quad
T^{({\rm EM})}_{ik} g^{ik}=0 \,.
\label{Ein7}
\end{equation}
The quantity $T^{({\rm EMA})}_{ik}$ is linear in the Maxwell tensor:
$$
T^{({\rm EMA})}_{ik} = \frac12 \Delta_{ik}  \left(\omega_1 F^{mn} + \omega_2 U^m F^{nq}U_q \right) \nabla_m U_n  +
$$
$$
+ \omega_1 F_{n (i} \nabla_{k)}U^n  - \frac12 \omega_2 U_i U_k   U^q F_{nq}U^l \nabla_l U^n  +
$$
\begin{equation}
+
\frac12 \nabla^m  \left\{\omega_1 U_{(i} F_{k)m} + \omega_2 U^n\left[U_iU_k F_{mn}  - U_m U_{(i} F_{k)n} \right]\right\} +
\frac12 \omega_1 U_i U_k  \nabla^m \left(F_{mn}U^n \right)
\,.
\label{Ein9}
\end{equation}
Thus, equations (\ref{E2}), (\ref{E9}) (with (\ref{E4}), (\ref{M209}), (\ref{M5})), equations (\ref{5A1}) (with (\ref{5A61}), (\ref{5A62}), (\ref{5A30-1})), and equations (\ref{Ein1}) (with (\ref{0Ein5}), (\ref{5Ein6}), (\ref{Ein9})) form the self-consistent set of master equations of the Einstein-Maxwell-aether model, corresponding to the second order of the effective field theory.

\section{SU(N) - generalization of the Einstein-Maxwell-aether theory: The formalism}

We construct the theory of the SU(N) - symmetric dynamic aether in analogy with the Einstein-Yang-Mills-Higgs theory. We stress, that in this paper we follow the definitions of the book \cite{Rubakov} (see Section 4.3.), and consider all the fields taking values in the Lie algebra of the gauge group SU(N) (adjoint representation). Let us describe main mathematical elements of this extended theory.

\subsection{Necessary elements of the SU(N) group theory}

For the reconstruction of the action functional, we take, first, the quantities ${\bf t}_{(a)}$, the Hermitian traceless generators of the SU(N) group; the group index $(a)$ runs from 1 to $N^2-1$.
The scalar product of the generators ${\bf t}_{(a)}$ and ${\bf t}_{(b)}$ is indicated as
\begin{equation}
\left( {\bf t}_{(a)} , {\bf t}_{(b)} \right) \equiv 2 {\rm Tr} \
{\bf t}_{(a)} {\bf t}_{(b)}  \equiv G_{(a)(b)}  \,.
\label{scalarproduct}
\end{equation}
The commutator of the generators
\begin{equation}
\left[ {\bf t}_{(a)} , {\bf t}_{(b)} \right] = i  f^{(c)}_{\
(a)(b)} {\bf t}_{(c)}  \label{fabc}
\end{equation}
introduces the structure constants of the gauge group SU(N), $f^{(c)}_{\ (a)(b)}$.
These structure constants satisfy the Jacobi identity
\begin{equation}
f^{(a)}_{\ (b)(c)} f^{(c)}_{\ (e)(h)}  + f^{(a)}_{\ (e)(c)} f^{(c)}_{\ (h)(b)} + f^{(a)}_{\ (h)(c)} f^{(c)}_{\ (b)(e)} = 0
\,. \label{jfabc}
\end{equation}
Also, one can introduce the completely symmetric coefficients
$d_{(c)(a)(b)}$ using the anti-commutator
\begin{equation}
\left\{ {\bf t}_{(a)} , {\bf t}_{(b)} \right\} \equiv {\bf
t}_{(a)}  {\bf t}_{(b)} + {\bf t}_{(b)} {\bf t}_{(a)} =
\frac{1}{N} \delta_{(a)(b)} {\bf I} + d^{(c)}_{\ (a)(b)} {\bf
t}_{(c)} \label{dabc}
\end{equation}
(see, e.g., \cite{Akhiezer}), where  ${\bf I}$ is the matrix-unity.

The symmetric tensor $G_{(a)(b)}$ plays the fundamental role in this theory, it is the metric in the group space. The generators can be chosen so that the metric is
equal to the Kronecker delta; in this sense, the metric introduces the universal tensor, since the structure of $G_{(a)(b)}$ depends on the dimension of the group space, but, in fact, it does not reveal features of SU(2), SU(3), SU(4),.. etc. With this metric tensor one can introduce the auxiliary quantities
\begin{equation}
f_{(c)(a)(b)} \equiv G_{(c)(d)} f^{(d)}_{\ (a)(b)} = - 2 i \
{\rm Tr} \ \left[ {\bf t}_{(a)} , {\bf t}_{(b)} \right] {\bf
t}_{(c)}  \,, \label{fabc1}
\end{equation}
and stress that $f_{(a)(b)(c)}$ are antisymmetric with respect to permutations of any two indices
\cite{Rubakov,Akhiezer}.
When the basis ${\bf t}_{(a)}$ is chosen to provide the relation
$G_{(a)(b)} = \delta_{(a)(b)}$, we obtain
\begin{equation}
\frac{1}{N} f^{(d)}_{\ (a)(c)} f^{(c)}_{\ (d)(b)} =
\delta_{(a)(b)} = G_{(a)(b)} \,.   \label{ff}
\end{equation}
There are also possibilities to introduce the tensors ${\cal F}_{(b)(c)(e)(f)}$ and ${\cal D}_{(a)(b)(e)(f)}$ in the group space, which are of the following form:
\begin{equation}\label{plus}
{\cal F}_{(b)(c)(e)(f)} \equiv f^{(a)}_{\ (b)(c)}f^{(d)}_{\ (e)(f)} G_{(a)(d)} \,,  \quad {\cal D}_{(a)(b)(e)(f)} \equiv d_{(a)(b)(c)} d_{(d)(e)(f)} G^{(c)(d)} \,.
\end{equation}
These tensors are not universal, they depend essentially on the specific features of the corresponding SU(N) group. Nevertheless, we mention them since, in principle, they can be used for reconstruction of constitutive tensors in some specific cases.

\subsection{Gauge field and field strength tensor}

We form the Yang-Mills field potential ${\bf A}_m$ and the Yang-Mills field strength ${\bf F}_{ik}$ as anti-Hermitian quantities:
\begin{equation}
{\bf A}_m = - i {\cal G} {\bf t}_{(a)} A^{(a)}_m \,, \quad {\bf F}_{mn} = - i {\cal G} {\bf t}_{(a)} F^{(a)}_{mn}  \,. \label{represent}
\end{equation}
The quantities $A^{(a)}_i$ and $F^{(a)}_{ik}$ describe two  multiplets of real fields, which are the SU(N) generalizations of the U(1) symmetric potential four-vector $A_i$ and of the Maxwell tensor $F_{ik}$, respectively.  They are connected by the well-known
formulas (see, e.g., \cite{Rubakov,Akhiezer})
\begin{equation}
{\bf F}_{mn} = \nabla_m {\bf A}_n - \nabla_n {\bf A}_m + \left[ {\bf
A}_m , {\bf A}_n \right]  \,, \label{45Fmn}
\end{equation}
\begin{equation}
F^{(a)}_{mn} = \nabla_m
A^{(a)}_n - \nabla_n A^{(a)}_m + {\cal G} f^{(a)}_{\ (b)(c)}
A^{(b)}_m A^{(c)}_n \,. \label{46Fmn}
\end{equation}
Here $\nabla _m$ is a  covariant space-time derivative.
Clearly, (\ref{46Fmn}) is the SU(N) generalization of (\ref{maxwellA1}).
The scalar invariant, which one uses in the Lagrangian of SU(N) symmetric theory is proportional to
\begin{equation}
I_1 \equiv \left({\bf F}_{mn},\,{\bf
F}^{mn}\right) \Rightarrow - {\cal G}^2 F^{(a)}_{mn}F_{(a)}^{mn}
\,.
\label{f2}
\end{equation}
Each element of the multiplet $F^{ik}_{(a)}$ has its own dual element defined as
\begin{equation}
{}^*\! F^{ik}_{(a)} = \frac{1}{2}\epsilon^{ikls} F_{ls (a)} \,,
\label{dual}
\end{equation}
with universal Levi-Civita tensor, introduced above.

\subsection{Vector field multiplet}

Extending the model of the  U(1)- symmetric aether, we introduce the multiplet of vector fields
\begin{equation}
{\bf U}_m =  {\bf t}_{(a)} U^{(a)}_m \,,  \label{Urepresent}
\end{equation}
in analogy with the Higgs multiplet of scalar fields in \cite{a1}
\begin{equation}
{\bf \Phi} = {\bf t}_{(a)} \Phi^{(a)} \,. \label{3represent}
\end{equation}
Thus, ${\bf U}_m $ and ${\bf \Phi}$ are considered to be Hermitian,
while ${\bf F}_{mn}$ and ${\bf A}_i$ are anti-Hermitian.
We assume that the trace of the vector fields is unit, or to be more precise, we assume that
\begin{equation}
G_{(a)(b)} U^{(a)}_m U^{(b)}_n g^{mn} = 1 \,. \label{4represent}
\end{equation}
Clearly, this condition is a generalization of the normalization condition $g^{mn}U_mU_n=1$ in the standard Einstein-aether theory.

\subsection{Gauge covariant derivatives}

In the extended theory of dynamic aether we use the operator $\hat{D}_m$, the extended (gauge covariant) derivative.  For the Higgs fields it is defined as (\cite{Rubakov}, Eqs.(4.46, 4.47))
$$
\D_m {\bf \Phi} \equiv \nabla_m {\bf \Phi} + \left[ {\bf A}_m ,
{\bf \Phi} \right] \,,
$$
\begin{equation}
\D_m \Phi^{(a)} \equiv \nabla_m
\Phi^{(a)} + {\cal G} f^{(a)}_{\ (b)(c)} A^{(b)}_m \Phi^{(c)}
\,. \label{DPhi}
\end{equation}
For the derivative of arbitrary tensor $Q^{(a) \cdot \cdot \cdot}_{\cdot \cdot \cdot (d)}$, defined in the group space,
we use the following rule \cite{Akhiezer}:
\begin{eqnarray}
\D_m Q^{(a) \cdot \cdot \cdot}_{\cdot \cdot \cdot (d)} \equiv
\nabla_m Q^{(a) \cdot \cdot \cdot}_{\cdot \cdot \cdot (d)} + {\cal
G} f^{(a)}_{\cdot (b)(c)} A^{(b)}_m Q^{(c) \cdot \cdot
\cdot}_{\cdot \cdot \cdot (d)} - {\cal G} f^{(c)}_{\cdot (b)(d)}
A^{(b)}_m Q^{(a) \cdot \cdot \cdot}_{\cdot \cdot \cdot (c)} +...
\,. \label{DQ2}
\end{eqnarray}
When we deal with the gauge covariant derivative of the vector fields, we use, respectively, the formula
\begin{equation}
\D_m U^{(a)}_n \equiv \nabla_m
U^{(a)}_n + {\cal G} f^{(a)}_{\ (b)(c)} A^{(b)}_m U^{(c)}_n
\,. \label{DU}
\end{equation}
The tensor $F^{ik}_{(a)}$ satisfies the relation
\begin{equation}
\hat{D}_k {}^*\! F^{ik}_{(a)} = 0 \,, \label{Aeq2}
\end{equation}
it is the generalization of (\ref{E9}).
The metric $G_{(a)(b)}$ and the
structure constants $f^{(d)}_{\ (a)(c)}$ are supposed to
be constant tensors in the standard and covariant manner
\cite{Akhiezer}. This means that
\begin{equation}
\partial_m G_{(a)(b)} = 0 \,, \quad \D_m G_{(a)(b)} = 0 \,, \qquad
\partial_m  f^{(a)}_{\ (b)(c)} = 0 \,, \quad \D_m
f^{(a)}_{\ (b)(c)} = 0 \,. \label{DfG}
\end{equation}
Finally, due to the relationship (\ref{4represent}), we obtain immediately
\begin{equation}
g^{pq} U^{(b)}_p \D_m U^{(a)}_q G_{(a)(b)} \equiv U^{(a)}_p \D_m U^p_{(a)}= \frac12 \D_m [U^{(a)}_p U^p_{(a)}] = 0
\,, \label{DU1}
\end{equation}
which is the generalization of the rule $U^k \nabla_m U_k = 0$ in the standard theory.

\section{A simple version of the Einstein-Yang-Mills-aether model}

We consider, first, the version of the theory, in which the unit dynamic vector field $U^i$ is unique and is not associated with the adjoint representation of the SU(N) group; we indicate this version of the theory as the Einstein-Yang-Mills-aether model. The corresponding action functional can be reconstructed as a simple generalization of (\ref{totalact}):
\begin{equation}\label{0SUNact}
S = \int d^4 x \sqrt{{-}g}\ \left\{ \frac{1}{2\kappa}\left[R {+} 2\Lambda {+} \lambda\left(g^{mn}U_{m}U_{n} {-}1 \right) {+} K^{ijmn} \nabla_i U_{m} \nabla_j
U_n \right] {+} \frac{1}{4} {\cal C}^{ikmn}_{(a)(b)}F^{(a)}_{ik}
F_{mn}^{(b)}  \right\}\,.
\end{equation}
In this model there are only two objects with single group index, namely, the multiplet of gauge potential four-vectors $A_{i}^{(a)}$,  and the gauge field strength tensor $F^{(a)}_{mn}$. The group space itself possesses the symmetric two-indices metric tensor $G_{(a)(b)}$, the antisymmetric set of three-indices group constants $f_{(a)(b)(c)}$, however, one can not construct the vector using these quantities.  This means that the group space is assumed to be isotropic, and we have to postulate that the term linear in $F^{(a)}_{mn}$ can not appear in the action functional in contrast to the U(1) - symmetric theory. The linear response tensor is now proportional to the metric in the group space $G_{(a)(b)}$, and to the modified U(1)-symmetric constitutive tensor $\tilde{C}^{ikmn}$ (see (\ref{M2b})):
$$
{\cal C}^{ikmn}_{(a)(b)}= G_{(a)(b)} \tilde{C}^{ikmn} =
$$
\begin{equation}
= \frac{1}{2\tilde{\mu}} G_{(a)(b)}\left[
g^{im} g^{kn}{-}g^{in} g^{km} {+} (\tilde{\varepsilon} \tilde{\mu} {-}1)\left(g^{im} U^kU^n {-} g^{in} U^kU^m {+}g^{kn} U^iU^m {-}g^{km} U^iU^n \right)\right] \,.
\label{eym}
\end{equation}
(Modifications contain replacements $\mu$ with $\tilde{\mu}$, and $\varepsilon$ with $\tilde{\varepsilon}$). The modifications of the model master equations are predictable; let us consider them briefly.

\subsection{The modified Yang-Mills equations}

The variation with respect to gauge potential $A^k_{(a)}$ gives the equations
\begin{equation}\label{0Col1}
\hat{D}_k H^{ik}_{(a)} = 0 \,, \quad  H^{ik}_{(a)} \equiv {\cal C}^{ikmn}_{(a)(b)} F^{(b)}_{mn} = \tilde{C}^{ikmn} F_{(a)mn} \,.
\end{equation}
As usual, this set of equations should be supplemented by
\begin{equation}\label{00Col1}
\hat{D}_k F^{*ik}_{(a)} = 0 \,.
\end{equation}
Keeping in mind the analogy with medium electrodynamics \cite{ME,HehlObukhov}, we can decompose the tensor  $F^{(a)}_{mn}$
using the four-vectors ${\cal E}^{(a)}_m$ and ${\cal B}^{(a)}_m$ (the SU(N)-symmetric analogs of the four-vectors of the electric field and magnetic excitation) as follows:
\begin{equation}\label{decomp2}
F^{(a)}_{mn} = {\cal E}^{(a)}_m U_n - {\cal E}^{(a)}_n U_m  - \epsilon _{mn}^{\ \ \ \ pq}{\cal B}^{(a)}_p U_q \,,
\end{equation}
\begin{equation}\label{decomp1}
{\cal E}^{(a)}_m \equiv F^{(a)}_{mn} U^n \,, \quad {\cal B}^{(a)}_m \equiv F^{*(a)}_{mn} U^n  \,.
\end{equation}
Similarly, we obtain the decomposition of the excitation tensor:
\begin{equation}\label{decomp21}
H_{(a)}^{mn} = {\cal D}_{(a)}^m U^n - {\cal D}_{(a)}^n U^m  - \epsilon ^{mn}_{\ \ \ \ pq}{\cal H}_{(a)}^p U^q \,,
\end{equation}
\begin{equation}\label{decomp11}
{\cal D}_{(a)}^m \equiv H_{(a)}^{mn} U_n \,, \quad {\cal H}_{(a)}^m \equiv H_{(a)}^{*mn} U^n  \,.
\end{equation}
The quantities ${\cal E}^{(a)}_m$, ${\cal B}^{(a)}_m$, ${\cal D}_{(a)}^m$  and ${\cal H}_{(a)}^m$ are linked by the following constitutive equations:
\begin{equation}\label{decomp111}
{\cal D}_{(a)}^m  = \tilde{\varepsilon} {\cal E}_{(a)}^m \,, \quad {\cal B}^{(a)}_m = \tilde{\mu} {\cal H}^{(a)}_m \,,
\end{equation}
thus providing the interpretation of the parameters $\tilde{\varepsilon}$ and $\tilde{\mu}$ as color analogs of the electromagnetic permittivities.

\subsection{Dynamic equations for the colored aether}

The variation of the total action functional with respect to the vector field yields
\begin{equation}
\nabla_m {\cal J}^{mn}_{({\rm A})}  = I^n_{({\rm A})} {+} \kappa I^n_{({\rm YM})} {+} \lambda \ U^n \,,
\label{YM1}
\end{equation}
where the terms ${\cal J}^{mn}_{({\rm A})}$ and $I^n_{({\rm A})}$ are already defined in (\ref{0A31}) and (\ref{0A4A}), respectively, and the new four-vector is
\begin{equation}
I^n_{({\rm YM)}} \equiv  \frac14 \left[\frac{\delta \tilde{C}^{ikpq}}{\delta U_n} \right] G_{(a)(b)}F^{(a)}_{ik}F^{(b)}_{pq} =
 \left(\tilde{\varepsilon} {-}\frac{1}{\tilde{\mu}} \right)  F^{kn}_{(a)}F^{(a)}_{km} U^m
\,. \label{A62}
\end{equation}
The Lagrange multiplier is also modified
\begin{equation}
\lambda = U_n \left[\nabla_m {\cal J}^{mn}_{({\rm A})} -
I^n_{({\rm A})} \right] - \kappa U_n I^n_{({\rm YM})} \,.
\label{A30-1}
\end{equation}

\subsection{Gravity field equations}

The result of variation of the total action functional with respect to
the metric $g^{ik}$ is
\begin{equation}
 R_{ik} {-} \frac{1}{2} R \ g_{ik} {-}  \Lambda g_{ik} =
 T^{({\rm U})}_{ik} + \kappa T^{({\rm YM})}_{ik} \,.
\label{Ein123}
\end{equation}
The term $T^{({\rm U})}_{ik}$ is already presented above by the formula
(\ref{0Ein5}).
The term  $T^{({\rm YM})}_{ik}$ is given by
$$
T^{({\rm YM})}_{ik} = \frac{1}{\tilde{\mu}} \left\{\left[\frac14
g_{ik}F^{(a)}_{mn}F^{mn}_{(a)}-F^{(a)}_{im}F_{k (a)}^{\ m}\right] + \right.
$$
\begin{equation}
\left. + (\tilde{\varepsilon} \tilde{\mu} -
1)U^p U^q\left[\left(\frac12 g_{ik}- U_iU_k\right) F^{m}_{\ \ p (a)}
F^{(a)}_{mq} - F^{(a)}_{ip}F_{kq(a)} \right] \right\}
\,. \label{Ein6}
\end{equation}
This stress-energy tensor is symmetric, $T^{({\rm YM})}_{ik}=T^{({\rm YM})}_{ki}$, and traceless, $g^{ik}T^{({\rm YM})}_{ik}=0$. It can be decomposed using three irreducible elements:
the energy density scalar $W^{(\rm YM)}$, energy flux four-vector ${\cal Q}^{(\rm YM)}_k$, and the pressure tensor ${\cal P}_{ik}^{(\rm YM)}$:
\begin{equation}
T^{({\rm YM})}_{ik} = W^{(\rm YM)} U_i U_k + U_i {\cal Q}^{(\rm YM)}_k  + U_k {\cal Q}^{(\rm YM)}_i + {\cal P}_{ik}^{(\rm YM)}\,,
\label{YMdec1}
\end{equation}
\begin{equation}
W^{(\rm YM)} \equiv U^i T^{({\rm YM})}_{ik} U^k = -\frac12 \left[\tilde{\varepsilon} {\cal E}_m^{(a)} {\cal E}^m_{(a)} + \frac{1}{\tilde{\mu}} {\cal B}_m^{(a)}{\cal B}_m^{(a)} \right] \,,
\label{YMdec2}
\end{equation}
\begin{equation}
{\cal Q}^{(\rm YM)}_k \equiv \Delta_k^m T^{({\rm YM})}_{mn} U^n = -\frac{1}{\tilde{\mu}} \epsilon_{kmpq}{\cal E}^m_{(a)}{\cal B}^{p(a)} U^q = - \epsilon_{kmpq}{\cal E}^m_{(a)}{\cal H}^{p(a)} U^q\,,
\label{YMdec3}
\end{equation}\begin{equation}
{\cal P}_{ik}^{(\rm YM)} \equiv \Delta_i^m  T^{({\rm YM})}_{mn} \Delta_k^n = \frac12 \Delta_{ik} \left[\tilde{\varepsilon} {\cal E}^m_{(a)}{\cal E}_m^{(a)} +
\frac{1}{\tilde{\mu}} {\cal B}^{m}_{(a)}{\cal B}_{m}^{(a)} \right] - \tilde{\varepsilon} {\cal E}_{i(a)}{\cal E}_{k}^{(a)} - \frac{1}{\tilde{\mu}}{\cal B}_{i(a)}{\cal B}_{k}^{(a)} \,.
\label{YMdec4}
\end{equation}
The four-vector ${\cal Q}^{(\rm YM)}_k $ is an analog of the Poynting vector in electrodynamics of continua \cite{ME}. The equations
(\ref{Ein123}), (\ref{0Col1}), (\ref{00Col1}) and  (\ref{YM1}) form the self-consistent system of equations for the Einstein-Yang-Mills-aether model.

\section{The full-format theory of the  SU(N)-symmetric aether}

\subsection{Action functional}

The action functional of the SU(N) extension of the Einstein-Maxwell-aether theory can be constructed using the following scheme: we take the functional (\ref{totalact}) and replace $U_m$ with $U^{(a)}_m$, $\nabla_m U_n$ with $\hat{D}_m U^{(a)}_{n}$, $F_{mn}$ with $F^{(a)}_{mn}$, $K^{ijmn}$ with ${\cal K}^{ijmn}_{(a)(b)}$, $A^{ikpq}$ with ${\cal A}^{[ik]pq}_{(a)(b)}$, $C^{ikmn}$ with ${\cal C}^{ikmn}_{(a)(b)}$. We obtain now
$$
S = \int d^4 x \sqrt{-g}\ \left\{ \frac{1}{2\kappa}\left[R + 2\Lambda + \lambda\left(g^{mn}U^{(a)}_{m}U^{(b)}_{n}G_{(a)(b)} -1 \right) + {\cal K}^{ijmn}_{(a)(b)}\hat{D}_i U^{(a)}_{m}\hat{D}_j
U^{(b)}_n \right]
+ \right.
$$
\begin{equation}\label{SUNact}
\left.
+\frac{1}{2} {\cal A}^{[ik]mn}_{(a)(b)}F^{(a)}_{ik} \hat{D}_m U^{(b)}_{n} + \frac{1}{4} {\cal C}^{ikmn}_{(a)(b)}F^{(a)}_{ik}
F_{mn}^{(b)}  \right\}\,.
\end{equation}
The constitutive tensors ${\cal K}^{ijmn}_{(a)(b)}$, ${\cal A}^{[ik]pq}_{(a)(b)}$ and  ${\cal C}^{ikmn}_{(a)(b)}$ have now two color indices, $(a)$ and $(b)$, and the structure of these constitutive tensors has to be discussed especially.
Master equations appear as a result of variation of the action functional (\ref{SUNact}) with respect to four quantities: the Lagrange multiplier $\lambda$, the vector fields $U^k_{(a)}$, the gauge potential four-vector $A_i^{(b)}$, and space-time metric $g^{ij}$. The variation with respect to $\lambda$ gives the normalization condition (\ref{4represent}). Other variation details are described in the next three subsections.

\subsection{Master equations for the gauge fields}

The variation procedure with respect to $A^{(a)}_i$ gives the equations, which we can standardly represent in the following form:
\begin{equation}\label{Col1}
\hat{D}_k H^{ik}_{(a)} = \Gamma^i_{(a)}\,, \quad H^{ik}_{(a)} = {\cal H}^{ik}_{(a)} + {\cal C}^{ikmn}_{(a)(b)} F^{(b)}_{mn}\,.
\end{equation}
Here
\begin{equation}\label{Col3}
{\cal H}^{ik}_{(a)} \equiv  {\cal A}^{[ik]mn}_{(a)(b)} \hat{D}_m U^{(b)}_n
\end{equation}
is the tensor of spontaneous induction, which does not depend on  $F^{(b)}_{mn}$ but is predetermined by the gauge covariant derivative of the vector fields $U^{(b)}_n$;
$H^{ik}_{(a)}$ is the total color excitation tensor linear in the Yang-Mills field strength $F^{(b)}_{mn}$.
The so-called color current $\Gamma^i_{(a)}$ is of the form:
\begin{equation}\label{Col4}
 \Gamma^i_{(a)} =   {\cal G} f^{(d)}_{\ (c)(a)} U^{(c)}_k \left[\frac{1}{\kappa}  {\cal K}^{imkn}_{(d)(b)} \hat{D}_m U^{(b)}_n +
 \frac12 {\cal A}^{[mn]ik}_{(b)(d)} F^{(b)}_{mn} \right]\,.
\end{equation}
These equations have to be supplemented by the equation (\ref{Aeq2}).

\subsection{Master equations for the vector fields $U^k_{(a)}$}

The variation procedure with respect to $U^j_{(a)}$ yields the set of equations in the standard form
\begin{equation}
\D_i {\cal J}^{ij}_{(a)}
 = \lambda \ U^j_{(a)}  +  {\cal I}^{j}_{(a)} \,,
\label{CU1}
\end{equation}
where we introduced the following definitions:
\begin{equation}
{\cal J}^{ij}_{(a)} = {\cal J}^{ij}_{(1)(a)}+ {\cal J}^{ij}_{(2)(a)} \,, \quad
{\cal I}^{j}_{(a)} = {\cal I}^{j}_{(1)(a)} + {\cal I}^{j}_{(2)(a)} + {\cal I}^{j}_{(3)(a)}\,,
\label{CU2}
\end{equation}
\begin{equation}
{\cal J}^{ij}_{(1)(a)} = {\cal K}^{imjn}_{(a)(b)} \hat{D}_m U^{(b)}_n \,, \quad  {\cal J}^{ij}_{(2)(a)} = \frac{1}{2} {\cal A}^{[mn]ij}_{(a)(b)}F^{(b)}_{mn} \,,
\label{CU3}
\end{equation}
\begin{equation}
{\cal I}^j_{(1)(a)} = \frac12 \left[\frac{\delta {\cal K}^{ikmn}_{(c)(b)}}{\delta U^{(a)}_j} \right]  \hat{D}_i U^{(c)}_m \hat{D}_k U^{(b)}_n  \equiv
\frac12 {\cal K}^{ikmnj}_{(c)(b)(a)}   \hat{D}_i U^{(c)}_m \hat{D}_k U^{(b)}_n\,,
\label{CU31}
\end{equation}
\begin{equation}
{\cal I}^j_{(2)(a)} = \frac{\kappa}{2} \left[\frac{\delta {\cal A}^{[ik]mn}_{(c)(b)}}{\delta U^{(a)}_j} \right]  F^{(c)}_{ik} \hat{D}_m U^{(b)}_n  \equiv
\frac{\kappa}{2}{\cal A}^{[ik]mnj}_{(c)(b)(a)} F^{(c)}_{ik} \hat{D}_m U^{(b)}_n  \,,
\label{CU32}
\end{equation}\begin{equation}
{\cal I}^j_{(3)(a)} = \frac{\kappa}{4} \left[\frac{\delta {\cal C}^{ikmn}_{(c)(b)}}{\delta U^{(a)}_j} \right]  F^{(c)}_{ik} F^{(b)}_{mn} \equiv
\frac{\kappa}{4} {\cal C}^{ikmnj}_{(c)(b)(a)} F^{(c)}_{ik} F^{(b)}_{mn}\,.
\label{CU33}
\end{equation}
In these terms the Lagrange multiplier has the standard form also:
\begin{equation}
\lambda =  \ U_j^{(a)} \left[\D_i {\cal J}^{ij}_{(a)}
- {\cal I}^j_{(a)} \right]  \,.
\label{CU4}
\end{equation}
When the constitutive tensors ${\cal K}^{ikmn}_{(c)(b)}$, ${\cal A}^{[ik]mn}_{(c)(b)}$ and ${\cal C}^{ikmn}_{(c)(b)}$ are reconstructed, the corresponding variational derivatives in (\ref{CU31}), (\ref{CU32}), (\ref{CU33}), can be calculated explicitly.

\subsection{Master equations for the gravitational field}

The gravitational field is described by the set of equations
\begin{equation}
 R_{pq} {-} \frac{1}{2} R  g_{pq} =  \Lambda g_{pq} - \lambda U^{(a)}_p  U^{(b)}_q G_{(a)(b)} +
 T^{(1)}_{pq} + \kappa \left[T^{(2)}_{pq}{+} T^{(3)}_{pq} \right] \,.
\label{CE1}
\end{equation}
The stress-energy tensor of the colored vector fields $T^{(1)}_{pq}$ is now of the form
$$
T^{(1)}_{pq} = \frac12 g_{pq} {\cal K}^{ijmn}_{(a)(b)}  \hat{D}_i U^{(a)}_m \hat{D}_j U^{(b)}_m - \left[\frac{\delta {\cal K}^{ij}_{(a)(b)mn}}{\delta g^{pq}} \right]  \hat{D}_i U^{m(a)} \hat{D}_j U^{n(b)} +
$$
\begin{equation}
+ G_{(a)(b)}\D^m \left[U^{(b)}_{(p}{\cal J}_{q)m}^{(1)(a)} -
{\cal J}_{m(p}^{(1)(a)}U^{(b)}_{q)} -
{\cal J}_{(pq)}^{(1)(a)} U^{(b)}_m \right] \,.
\label{CE2}
\end{equation}
The tensor $T^{(2)}_{ik}$ describing the interaction between the gauge and vector fields contains the terms:
$$
T^{(2)}_{pq} = \frac12 g_{pq} {\cal A}^{[ik]mn}_{(a)(b)}F^{(a)}_{ik} \hat{D}_m U^{(b)}_{n} - \left[\frac{\delta {\cal A}^{[ik]m}_{(a)(b)n}}{\delta g^{pq}} \right]F^{(a)}_{ik} \hat{D}_m U^{(b) n} +
$$
\begin{equation}
 + G_{(a)(b)}\D^m \left[U^{(b)}_{(p}{\cal J}_{q)m}^{(2)(a)} -
{\cal J}_{m(p}^{(2)(a)}U^{(b)}_{q)} -
{\cal J}_{(pq)}^{(2)(a)} U^{(b)}_m \right] \,.
\label{CE3}
\end{equation}
The stress-energy tensor of the Yang-Mills field $T^{(3)}_{ik}$ can be presented as follows:
\begin{equation}
T^{(3)}_{pq} = \frac14 g_{pq}  {\cal C}^{ikmn}_{(a)(b)} F^{(a)}_{ik} F^{(b)}_{mn} - \frac{1}{2} \left[\frac{\delta {\cal C}^{ikmn}_{(a)(b)}}{\delta g^{pq}} \right]  F^{(a)}_{ik} F^{(b)}_{mn}\,.
\label{CE4}
\end{equation}
Variational derivatives in (\ref{CE2}), (\ref{CE3}) and (\ref{CE4}) can be calculated, when the constitutive tensors ${\cal K}^{ijmn}_{(a)(b)}$, ${\cal A}^{[ik]mn}_{(a)(b)}$ and
${\cal C}^{ikmn}_{(a)(b)}$ are presented explicitly; we reconstruct these quantities in the next section.

\subsection{Reconstruction of the constitutive tensors}

\subsubsection{The structure of the constitutive tensor ${\cal K}^{ijmn}_{(a)(b)}$}

We assume that the tensor ${\cal K}^{ijmn}_{(a)(b)}$ can be reconstructed using the space-time metric $g^{ik}$, the metric in the group space $G_{(a)(b)}$, the vector fields $U^{(a)}_m$, and
an appropriate number of coupling constants $C_n$. Since the tensor ${\cal K}^{ijmn}_{(a)(b)}$ appears in the Lagrangian in front of the quadratic combination of the gauge covariant derivative of the vector field, we have to keep in mind that this object possesses the following symmetry of indices:
\begin{equation}\label{9H9}
{\cal K}^{ijmn}_{(a)(b)} = {\cal K}^{jinm}_{(b)(a)}\,.
\end{equation}
We extend the decomposition, proposed by Jacobson and colleagues, as follows: first, we list all the terms, in which the space-time indices are provided by the product of the metric coefficients ($g^{ij}g^{mn}$, etc...); second, we list all the possible terms, which contain the product of the metric coefficients and two vector fields ($g^{ij}U^{m}_{(a)} U^n_{(b)}$, etc...); third, we list all the terms, in which the space-time indices are provided by the product of four vector fields ($U^i_{(a)}U^j_{(b)}U^m_{(c)}U^n_{(d)}$, etc...). Clearly, in the U(1)-symmetric theory the product $U^i U^j U^m U^n$ could not appear, however, the presence of the group indices changes the situation essentially.

In order to visualize the ways, along with the pairs of group indices $(a)(b)$ appear in the constitutive tensors, we introduce (in addition to the metric tensor $G_{(a)(b)}$)  the auxiliary tensor in the group space, $H^{(a)(b)}$:
\begin{equation}\label{H}
H^{(a)(b)} \equiv g^{pq} U_p^{(a)} U_q^{(b)}\,.
\end{equation}
The tensor $H^{(a)(b)}$ is associated with the multiplet of the vector fields, and it can be indicated as an analog of the polarization tensor in the optics (see, e.g., \cite{LL2}). In fact, it is the tensor of color polarization, it is real and symmetric. The matrix $H^{(a)}_{(b)}$ associated with this symmetric tensor  has $N^2{-}1$ eigenvalues,
$\sigma_{\{\alpha\}}$, and the corresponding $N^2{-}1$ eigenvectors $q^{(a)}_{\{\alpha\}}$:
\begin{equation}\label{H93}
H^{(a)}_{(b)}q^{(b)}_{\{\alpha\}}  = \sigma_{\{\alpha\}} q^{(a)}_{\{\alpha\}} \,.
\end{equation}
The sum of eigenvalues is equal to one, since the trace is equal to one due to the relationship
\begin{equation}\label{H90}
H^{(a)}_{(a)} = G_{(a)(b)}g^{mn} U^{(a)}_m U^{(b)}_n = 1 \,.
\end{equation}
One can decompose the tensor $H^{(a)(b)}$ in the series of products of eigenvectors:
\begin{equation}\label{H97}
H^{(a)(b)} = \sum_{\{\alpha\}} \sigma_{\{\alpha\}}q^{(a)}_{\{\alpha\}}q^{(b)}_{\{\alpha\}}  \,.
\end{equation}
In the first limiting case, when all the eigenvalues are equal to one another, and thus  $\sigma_{\{\alpha\}} = \sigma = 1/(N^2-1)$, we deal with the color analog of the so-called natural light \cite{LL2}. In the second limiting case, when all the eigenvalues, except one, are equal to zero, we deal with the color analog of the linearly polarized light. In the last case only one eigenvalue is non-vanishing, and it is equal to one; we obtain now $H^{(a)(b)}= q^{(a)}q^{(b)}$, where $q^{(a)}$ is the corresponding eigenvector.
Keeping in mind the theory of Stokes parameters in the optics, based on the analysis of the two-dimensional polarization tensor \cite{LL2}, one can develop the theory of "color Stokes parameters" based on the analysis of the color polarization tensor  $H_{(a)(b)}$  (it is a good idea for the future work).

All the contributions to the tensor ${\cal K}^{ijmn}_{(a)(b)}$, which are constructed based on the product of  space-time metric tensors, can be presented using six coupling constants
\begin{equation}\label{1D}
{}_{(1)}{\cal K}^{ijmn}_{(a)(b)} {=} G_{(a)(b)} \left[C_1 g^{ij} g^{mn} {+} C_2 g^{im}g^{jn}
{+} C_3 g^{in}g^{jm} \right] {+}  H_{(a)(b)} \left[C_{5} g^{ij} g^{mn} {+} C_{6} g^{im}g^{jn}
{+} C_{7} g^{in}g^{jm} \right].
\end{equation}
The parameters $C_1$, $C_2$ and $C_3$ are, in fact, extracted from the Jacobson's decomposition in the standard Einstein-aether theory; $C_5$, $C_6$ and $C_7$ appear in the SU(N) - extended theory only.

Contributions of the second type, based on the products of the space-time metric and two vector fields, can be listed as follows:
$$
{}_{(2)}{\cal K}^{ijmn}_{(a)(b)} = C_{4} U^{i}_{(a)} U^{j}_{(b)} g^{mn} + C_{8} U^{n}_{(a)} U^{m}_{(b)}g^{ij} +
C_{9} \left[U^{n}_{(a)} U^{j}_{(b)}g^{im} + U^{m}_{(b)} U^{i}_{(a)}g^{jn} \right] +
$$
$$
+ G_{(a)(b)}G^{(c)(d)}\left\{C_{10} U^{i}_{(c)} U^{j}_{(d)}g^{mn} + C_{11} U^{n}_{(c)} U^{m}_{(d)}g^{ij} +
\right.
$$
$$
\left. + C_{12} \left[ U^{n}_{(c)} U^{j}_{(d)}g^{im} + U^{i}_{(c)} U^{m}_{(d)}g^{jn} \right] +
C_{13} \left[ U^{m}_{(c)} U^{j}_{(d)}g^{in} + U^{i}_{(c)} U^{n}_{(d)}g^{jm} \right] \right\}
+
$$
$$
+ G_{(a)(b)}H^{(c)(d)}\left\{C_{14} U^{i}_{(c)} U^{j}_{(d)}g^{mn} + C_{15} U^{n}_{(c)} U^{m}_{(d)}g^{ij} +
\right.
$$
$$
\left. + C_{16} \left[ U^{n}_{(c)} U^{j}_{(d)}g^{im} + U^{i}_{(c)} U^{m}_{(d)}g^{jn} \right] +
C_{17} \left[ U^{m}_{(c)} U^{j}_{(d)}g^{in} + U^{i}_{(c)} U^{n}_{(d)}g^{jm} \right] \right\}
+
$$
$$
+ H_{(a)(b)}G^{(c)(d)}\left\{C_{18} U^{i}_{(c)} U^{j}_{(d)}g^{mn} + C_{19} U^{n}_{(c)} U^{m}_{(d)}g^{ij} +
\right.
$$
$$
\left. + C_{20} \left[ U^{n}_{(c)} U^{j}_{(d)}g^{im} + U^{i}_{(c)} U^{m}_{(d)}g^{jn} \right] +
C_{21} \left[ U^{m}_{(c)} U^{j}_{(d)}g^{in} + U^{i}_{(c)} U^{n}_{(d)}g^{jm} \right] \right\}
+
$$
$$
+ H_{(a)(b)}H^{(c)(d)}\left\{C_{22} U^{i}_{(c)} U^{j}_{(d)}g^{mn} + C_{23} U^{n}_{(c)} U^{m}_{(d)}g^{ij} +
\right.
$$
\begin{equation}\label{51D}
\left. + C_{24} \left[ U^{n}_{(c)} U^{j}_{(d)}g^{im} + U^{i}_{(c)} U^{m}_{(d)}g^{jn} \right] +
C_{25} \left[ U^{m}_{(c)} U^{j}_{(d)}g^{in} + U^{i}_{(c)} U^{n}_{(d)}g^{jm} \right] \right\}
\,.
\end{equation}
Thus, in addition to the Jacobson's constant $C_4$, eighteen new parameters appear in the SU(N) - extended theory. By the way, in the U(1) - symmetric theory, the terms with the coefficients $U^{n}U^{m}g^{ij}$ disappear automatically because of the normalization condition (see Remark), nevertheless, in the SU(N) - extended theory the corresponding  terms with $U^{n}_{(a)} U^{m}_{(b)}g^{ij}$ give non-vanishing expressions in general case.

Contributions of the third type, in which the product of four vector fields provides the presence of free indices $ijmn$, is absolutely new: these terms can not appear in the U(1) - symmetric theory because of the normalization condition. Sixteen coupling constants are included into the decomposition:
$$
{}_{(3)}{\cal K}^{ijmn}_{(a)(b)} =  G^{(c)(d)}\left\{
C_{26} U^{i}_{(c)}U^{j}_{(d)}U^{n}_{(a)}U^{m}_{(b)}+
C_{27} \left[U^{i}_{(c)}U^{m}_{(d)} U^{n}_{(a)}U^{j}_{(b)} + U^{j}_{(c)}U^{n}_{(d)} U^{i}_{(a)}U^{m}_{(b)} \right] \right. +
$$
$$
\left.
+C_{28}\left[U^{i}_{(c)}U^{n}_{(d)}U^{j}_{(a)}U^{m}_{(b)} + U^{j}_{(c)}U^{m}_{(d)}U^{i}_{(b)}U^{n}_{(a)}\right]
+
U^{m}_{(c)}U^{n}_{(d)}\left[C_{29}U^{i}_{(a)}U^{j}_{(b)} + C_{30}U^{j}_{(a)}U^{i}_{(b)}\right] \right\} +
$$
$$
+ H^{(c)(d)}\left\{
C_{31} U^{i}_{(c)}U^{j}_{(d)}U^{n}_{(a)}U^{m}_{(b)}+
C_{32} \left[U^{i}_{(c)}U^{m}_{(d)} U^{n}_{(a)}U^{j}_{(b)} + U^{j}_{(c)}U^{n}_{(d)} U^{i}_{(a)}U^{m}_{(b)} \right] \right. +
$$
$$
\left.
+C_{33}\left[U^{i}_{(c)}U^{n}_{(d)}U^{j}_{(a)}U^{m}_{(b)} + U^{j}_{(c)}U^{m}_{(d)}U^{i}_{(b)}U^{n}_{(a)}\right]
+
U^{m}_{(c)}U^{n}_{(d)}\left[C_{34}U^{i}_{(a)}U^{j}_{(b)} + C_{35}U^{j}_{(a)}U^{i}_{(b)}\right] \right\} +
$$
$$
+ \left\{\left[C_{36}G_{(a)(b)}+ C_{37} H_{(a)(b)}\right]\left[G^{(e)(f)}G^{(c)(d)}+G^{(e)(c)}G^{(f)(d)}+ G^{(e)(d)}G^{(c)(f)}\right]+ \right.
$$
$$
\left.
+ \left[C_{38} G_{(a)(b)} + C_{39}H_{(a)(b)} \right]\left[ H^{(e)(f)} G^{(c)(d)}+ H^{(e)(c)} G^{(f)(d)} + H^{(e)(d)} G^{(c)(f)}+  \right. \right.
$$
$$
\left. \left.  + H^{(c)(d)} G^{(e)(f)} + H^{(c)(f)} G^{(e)(d)} + H^{(f)(d)} G^{(e)(c)} \right] + \right.
$$
$$
\left.
+\left[C_{40}G_{(a)(b)}+ C_{41} H_{(a)(b)}\right] \left[ H^{(e)(f)} H^{(c)(d)}  + H^{(e)(c)} H^{(f)(d)} + H^{(e)(d)} H^{(c)(f)}\right]\right\} \times
$$
\begin{equation}
\times {\cal P}\left[U^{i}_{(e)}U^{m}_{(f)}U^{j}_{(c)}U^{n}_{(d)}\right]
\,.
\label{q2}
\end{equation}
The term ${\cal P}\left[U^{i}_{(e)}U^{m}_{(f)}U^{j}_{(c)}U^{n}_{(d)}\right]$ in the last line of this decomposition denotes the whole set of combinatoric permutations of the  elements $U^{i}_{(e)}U^{m}_{(f)}U^{j}_{(c)}U^{n}_{(d)}$ with respect to space-time indices; it contains the sum of 24 items $U^{i}_{(e)}U^{m}_{(f)}U^{j}_{(c)}U^{n}_{(d)}+ U^{i}_{(e)}U^{m}_{(f)}U^{n}_{(c)}U^{j}_{(d)}+ ...$. Let us emphasize that in front of the term ${\cal P}\left[U^{i}_{(e)}U^{m}_{(f)}U^{j}_{(c)}U^{n}_{(d)}\right]$ we use the tensor in the group space, which is absolutely symmetric with respect to group indices $(e)(f)(c)(d)$; if the Reader prefers to break this symmetry, one can enlarge this decomposition thus increasing the number of coupling constants.

\subsubsection{The structure of the constitutive tensor ${\cal A}^{[ik]pq}_{(a)(b)}$}

The constitutive tensor ${\cal A}^{[ik]mn}_{(a)(b)}$ can be reconstructed using the procedure similar to the one applied for the tensor ${\cal K}^{ijmn}_{(a)(b)}$.
Again, we consider, first, the terms in which the product of two metric tensors provide the presence of four space-time indices; second, we list all the terms based on the products of one metric tensor and two vector fields; third, we introduce the terms proportional to the product of four vector fields (such terms are absolutely new in comparison with U(1) symmetric theory). Following this procedure we obtain
$$
{\cal A}^{[ik]mn}_{(a)(b)} = \left[\omega_1 G_{(a)(b)}+ \omega_2 H_{(a)(b)} \right] g^{m[i} g^{k]n} +
$$
$$
+ \omega_3 U^{[i}_{(a)} U^{k]}_{(b)} g^{mn} + \omega_4  g^{n[i} U^{k]}_{(a)} U^{m}_{(b)} + \omega_5  g^{n[i} U^{k]}_{(b)} U^{m}_{(a)} + \omega_6  g^{m[i} U^{k]}_{(b)} U^{n}_{(a)}
+
$$
$$
+ G_{(a)(b)} G^{(c)(d)}\left[\omega_7  g^{n[i} U^{k]}_{(c)} U^{m}_{(d)} +  \omega_8  g^{m[i} U^{k]}_{(c)} U^{n}_{(d)} \right] +
$$
$$
+ H_{(a)(b)} G^{(c)(d)}\left[\omega_{9}  g^{n[i} U^{k]}_{(c)} U^{m}_{(d)} +  \omega_{10}  g^{m[i} U^{k]}_{(c)} U^{n}_{(d)} \right] +
$$
$$
+ G_{(a)(b)} H^{(c)(d)}\left[\omega_{11}  g^{n[i} U^{k]}_{(c)} U^{m}_{(d)} +  \omega_{12}  g^{m[i} U^{k]}_{(c)} U^{n}_{(d)} \right] +
$$
$$
+ H_{(a)(b)} H^{(c)(d)}\left[\omega_{13}  g^{n[i} U^{k]}_{(c)} U^{m}_{(d)} +  \omega_{14}  g^{m[i} U^{k]}_{(c)} U^{n}_{(d)} \right] +
$$
$$
{+}G^{(c)(d)}\left\{\omega_{15} U^{[i}_{(a)} U^{k]}_{(b)} U^{m}_{(c)} U^{n}_{(d)} +
\omega_{16} U^{[i}_{(a)} U^{k]}_{(c)} U^{m}_{(b)} U^{n}_{(d)} +
U^{[i}_{(b)} U^{k]}_{(c)} \left[\omega_{17}  U^{m}_{(a)} U^{n}_{(d)} +
\omega_{18} U^{m}_{(d)} U^{n}_{(a)}\right] \right\} {+}
$$
\begin{equation}\label{A1}
{+}H^{(c)(d)}\left\{\omega_{19} U^{[i}_{(a)} U^{k]}_{(b)} U^{m}_{(c)} U^{n}_{(d)} {+}
\omega_{20} U^{[i}_{(a)} U^{k]}_{(c)} U^{m}_{(b)} U^{n}_{(d)} {+}
U^{[i}_{(b)} U^{k]}_{(c)} \left[\omega_{21}  U^{m}_{(a)} U^{n}_{(d)} {+}
\omega_{22} U^{m}_{(d)} U^{n}_{(a)}\right] \right\} \,.
\end{equation}
Let us mention that the terms of the type  ${\cal P}\left[U^{[i}_{(e)} U^{k]}_{(f)} U^{m}_{(c)} U^{n}_{(d)}\right]$ with skew-symmetric set of indices $ik$ disappear, since we assume that the dumb indices $(e)(f)(c)(d)$ form  absolutely symmetric set, similarly to the case considered above for the tensor ${}_{(3)}{\cal K}^{ijmn}_{(a)(b)}$.

\subsubsection{The structure of the constitutive tensor ${\cal C}^{ikmn}_{(a)(b)}$}

In order to reconstruct the quantity ${\cal C}^{ikmn}_{(a)(b)}$, the SU(N) analog of the linear response tensor, we take the U(1)-symmetric constitutive tensor (\ref{M2b}) and list all the irreducible terms quadratic in the space-time metric, quadratic and quartic in the vector field, respectively. We take into account the symmetry of this tensor
\begin{equation}
{\cal C}^{ikmn}_{(a)(b)}=-{\cal C}^{kimn}_{(a)(b)}=-{\cal C}^{iknm}_{(a)(b)}={\cal C}^{mnik}_{(b)(a)} \,
\label{symmetryC}
\end{equation}
The result of decomposition is the following:
$$
{\cal C}^{ikmn}_{(a)(b)} = \left[\Omega_1 G_{(a)(b)} + \Omega_2 H_{(a)(b)} \right]\left[g^{im} g^{kn} - g^{in}g^{km}\right] +
$$
$$
+ \Omega_3 \left[g^{im} U^k_{(a)}U^n_{(b)} - g^{in} U^k_{(a)} U^m_{(b)} +g^{kn} U^i_{(a)} U^m_{(b)} -g^{km} U^i_{(a)}U^n_{(b)}  \right]+
$$
$$
+ \left\{\left[\Omega_4 G_{(a)(b)} + \Omega_5 H_{(a)(b)}\right] G^{(c)(d)} + \left[\Omega_6 G_{(a)(b)}+ \Omega_7 H_{(a)(b)} \right]H^{(c)(d)} + \right.
$$
$$
\left. +\Omega_8 \left[H_{(a)}^{\ (c)} G_{(b)}^{\ (d)} + H_{(a)}^{\ (d)} G_{(b)}^{\ (c)} + H_{(b)}^{\ (d)} G_{(a)}^{\ (c)} + H_{(b)}^{\ (c)} G_{(a)}^{\ (d)} \right]  \right\}\times
$$
$$
\times
\left[g^{im} U^k_{(c)}U^n_{(d)} - g^{in} U^k_{(c)} U^m_{(d)} +g^{kn} U^i_{(c)} U^m_{(d)} -g^{km} U^i_{(c)}U^n_{(d)}  \right] +
$$
\begin{equation}\label{vA1}
+ \left[\Omega_9 G^{(c)(d)} + \Omega_{10} H^{(c)(d)} \right] U^{[i}_{(a)} U^{k]}_{(c)}U^{[m}_{(b)}U^{n]}_{(d)} + \left[\Omega_{11} G^{(c)(d)} + \Omega_{12} H^{(c)(d)} \right] U^{[i}_{(b)} U^{k]}_{(c)}U^{[m}_{(a)}U^{n]}_{(d)} \,.
\end{equation}
In (\ref{M2b}) there are only two coupling parameters, $\varepsilon$ and $\mu$, while in (\ref{vA1}) we see twelve coupling constants, $\Omega_1, ... \Omega_{12}$.

\section{Models with spontaneous color polarization}

\subsection{The ansatz}

In order to reduce the number of coupling constants (the number $75=41+22+12$ for guiding parameters of the model seems to be too large) we have to formulate some simplifying assumptions. One of the ideas is to consider the multiplet of the vector fields to be "parallel" in the group space
\begin{equation}
U^i_{(a)}= q_{(a)}U^i  \,,
\label{ans1}
\end{equation}
where the fundamental vector field $U^i$ is unit and time-like
\begin{equation}
g_{ik} U^i U^k =1 \,,
\label{ans11}
\end{equation}
and the vector in the group space $q^{(a)}$ is also unit
\begin{equation}
G_{(a)(b)}q^{(a)} q^{(b)} = 1 \,.
\label{ans111}
\end{equation}
This  normalization condition means that the gauge-covariant derivative $\D_m q^{(a)}$ is orthogonal to the color vector $q_{(a)}$, i.e.,
\begin{equation}
q_{(a)}\D_m q^{(a)} = 0  \ \Rightarrow  q_{(a)} \left[\partial_m q^{(a)} + {\cal G} f^{(a)}_{\ \ (b)(c)} A^{b}_m q^{(c)}\right] =0 \Rightarrow  q_{(a)} \partial_m q^{(a)} =0\,.
\label{1ans1}
\end{equation}
Also, one can see that
\begin{equation}
\D_m U^{(a)}_n = q^{(a)}\nabla_m U_n + U_n \D_m q^{(a)}   \,.
\label{2ans1}
\end{equation}
The auxiliary tensor $H_{(a)(b)}$ converts now into
\begin{equation}
H_{(a)(b)}= q_{(a)} q_{(b)} \,,
\label{ans109}
\end{equation}
and the color vector $q_{(a)}$ is an eigen-vector of $H_{(a)(b)}$ with unit eigen-value
\begin{equation}
H_{(a)(b)} q^{(a)} = q_{(b)}\,.
\label{ans109b}
\end{equation}
This ansatz introduces a special direction in the group space, which is pointed by the color vector $q_{(a)}$, and return us to the global unit vector field $U^i$.
When the so-called "color poling" is implemented, and all the vector fields $U^i_{(a)}$ are oriented along the vector $q_{(a)}$ in the group space (it could be indicated as a phase transition), the tensors ${\cal K}^{ijmn}_{(a)(b)}$, ${\cal A}^{[ik]pq}_{(a)(b)}$ and ${\cal C}^{ikmn}_{(a)(b)}$ are simplified essentially.

\subsection{Reduced tensor ${\cal K}^{ijmn}_{(a)(b)}$}

With the ansatz (\ref{ans1}) the elements of the first constitutive tensor ${\cal K}^{ijmn}_{(a)(b)}$ take the form:
$$
{}_{(1)}{\cal K}^{ijmn}_{(a)(b)} = G_{(a)(b)} \left[C_1 g^{ij} g^{mn} {+} C_2 g^{im}g^{jn}
{+} C_3 g^{in}g^{jm} \right] {+}
$$
$$
+ q_{(a)} q_{(b)} \left[C_{5} g^{ij} g^{mn} {+} C_{6} g^{im}g^{jn}
{+} C_{7} g^{in}g^{jm} \right] \,,
$$

$$
{}_{(2)}{\cal K}^{ijmn}_{(a)(b)} =  U^{i} U^{j}g^{mn} \left[\left(C_{4}+ C_{18}+ C_{22} \right) q_{(a)} q_{(b)}  +
G_{(a)(b)}\left( C_{10}  +  C_{14}\right) \right] \,,
$$
\begin{equation}
{}_{(3)}{\cal K}^{ijmn}_{(a)(b)} =  0 \,.
\label{1D9}
\end{equation}
Let us introduce the color projector, the auxiliary tensor in the group space, as follows:
\begin{equation}
\Pi_{(a)(b)} \equiv G_{(a)(b)} - q_{(a)} q_{(b)}\,.
\label{P1}
\end{equation}
This color projector possesses the properties:
\begin{equation}
\Pi_{(a)(b)} q^{(b)} = 0 \,, \quad \Pi_{(a)(b)} = \Pi_{(b)(a)}  \,, \quad G^{(a)(b)}  \Pi_{(a)(b)} = N^2-2 \,.
\label{P2}
\end{equation}
Then the tensor ${\cal K}^{ijmn}_{(a)(b)}$ splits into the longitudinal and transversal parts:
\begin{equation}
{\cal K}^{ijmn}_{(a)(b)} = q_{(a)}q_{(b)} {\cal K}^{ijmn}_{(\rm long)} +
\Pi_{(a)(b)} {\cal K}^{ijmn}_{(\rm trans)}   \,,
\label{P3}
\end{equation}
where
$$
{\cal K}^{ijmn}_{(\rm long)} =  (C_1{+}C_5) g^{ij} g^{mn} {+} (C_2{+}C_6) g^{im}g^{jn}
{+} (C_3{+}C_7) g^{in}g^{jm} {+}
$$
\begin{equation}
{+}(C_4 {+} C_{10}{+}C_{14} {+} C_{18}{+} C_{22}) U^{i} U^{j}g^{mn}   \,,
\label{P4}
\end{equation}
\begin{equation}
{\cal K}^{ijmn}_{(\rm trans)} = C_1 g^{ij} g^{mn} {+} C_2 g^{im}g^{jn}
{+} C_3 g^{in}g^{jm} {+} (C_{10}{+}C_{14}) U^{i} U^{j}g^{mn}  \,.
\label{P5}
\end{equation}
Clearly, both longitudinal and transversal constitutive tensors have the same structure as the Jacobson ones (\ref{2}), the difference is in the coupling constants only. In other words,
we can rewrite longitudinal and transversal constitutive tensors as
\begin{equation}
{\cal K}^{ijmn}_{(\rm long)} =  C^{(||)}_1 g^{ij} g^{mn} {+} C^{(||)}_2 g^{im}g^{jn}
{+} C^{(||)}_3 g^{in}g^{jm} {+}C^{(||)}_4 U^{i} U^{j}g^{mn}   \,,
\label{P47}
\end{equation}
\begin{equation}
{\cal K}^{ijmn}_{(\rm trans)} = C_1 g^{ij} g^{mn} {+} C_2 g^{im}g^{jn}
{+} C_3 g^{in}g^{jm} {+} C^{(\bot)}_{4} U^{i} U^{j}g^{mn}  \,,
\label{P57}
\end{equation}
where
$$
C^{(||)}_1 \equiv C_1{+}C_5 \,, \quad  C^{(||)}_2 \equiv C_2{+}C_6 \,, \quad C^{(||)}_3 \equiv C_3{+}C_7 \,,
$$
\begin{equation}
C^{(||)}_4 \equiv C_4 {+} C_{10}{+}C_{14} {+} C_{18}{+} C_{22} \,, \quad
C^{(\bot)}_{4} \equiv C_{10}{+}C_{14}
\label{P571}
\end{equation}
present five effective coupling constants, additional to the standard set $C_1$, $C_2$ and $C_3$.

\subsection{Reduced tensor ${\cal A}^{[ik]mn}_{(a)(b)}$}

Similarly to the first case, one can obtain the reduced version of the second constitutive tensor ${\cal A}^{[ik]mn}_{(a)(b)}$; it has the form
$$
{\cal A}^{[ik]mn}_{(a)(b)} = \left[\omega_1 G_{(a)(b)}+ \omega_2 q_{(a)} q_{(b)} \right] g^{m[i} g^{k]n} +
$$
\begin{equation}\label{A11}
+ \left[ G_{(a)(b)} (\omega_7+ \omega_{11})  +
q_{(a)} q_{(b)} (\omega_4 +\omega_5+\omega_{9}+ \omega_{13})  \right] g^{n[i} U^{k]} U^{m}
\,,
\end{equation}
and can be rewritten as follows:
\begin{equation}\label{A12}
{\cal A}^{[ik]mn}_{(a)(b)} = q_{(a)} q_{(b)} {\cal A}^{[ik]mn}_{(\rm long)} + \Pi_{(a)(b)}{\cal A}^{[ik]mn}_{(\rm trans)}
\,,
\end{equation}
\begin{equation}\label{A13}
{\cal A}^{[ik]mn}_{(\rm long)} \equiv \left(\omega_1 {+} \omega_2 \right) g^{m[i} g^{k]n}
{+} \left(\omega_4 {+} \omega_5 {+} \omega_7 {+} \omega_{9} {+} \omega_{11}
{+} \omega_{13} \right) g^{n[i} U^{k]} U^{m}
\,,
\end{equation}
\begin{equation}\label{A14}
{\cal A}^{[ik]mn}_{(\rm trans)} \equiv \omega_1 g^{m[i} g^{k]n} + (\omega_7+ \omega_{11}) g^{n[i} U^{k]} U^{m}
\,.
\end{equation}

\subsection{Reduced tensor ${\cal C}^{ikmn}_{(a)(b)}$}

Using the ansatz about color polarization, one can reduce the linear response tensor as follows:
$$
{\cal C}^{ikmn}_{(a)(b)} = \left[\Omega_1 \Pi_{(a)(b)} + (\Omega_1 +\Omega_2) q_{(a)} q_{(b)} \right]\left[g^{im} g^{kn} - g^{in}g^{km}\right] +
$$
$$
+ \left[(\Omega_4+\Omega_6) \Pi_{(a)(b)} + (\Omega_3+\Omega_4+\Omega_5+\Omega_6 +\Omega_7+ 4 \Omega_8) q_{(a)} q_{(b)} \right] \times
$$
\begin{equation}\label{v09}
\times \left[g^{im} U^k U^n - g^{in} U^k U^m + g^{kn} U^i U^m  - g^{km} U^i U^n  \right] \,.
\end{equation}
Again, this tensor splits into the longitudinal and transversal parts
\begin{equation}\label{9}
{\cal C}^{ikmn}_{(a)(b)} = q_{(a)} q_{(b)} {\cal C}^{ikmn}_{(\rm long)} + \Pi_{(a)(b)} {\cal C}^{ikmn}_{(\rm trans)}
\,,
\end{equation}
\begin{equation}\label{m}
{\cal C}^{ikmn}_{(\rm long)}  = \frac{1}{2\mu_{(||)}}\left\{g^{ikmn} + \left[\varepsilon_{(||)} \mu_{(||)} {-}1 \right]
\left[g^{im} U^k U^n {-} g^{in} U^k U^m {+} g^{kn} U^i U^m  {-} g^{km} U^i U^n  \right]\right\} \,,
\end{equation}
\begin{equation}\label{mn}
{\cal C}^{ikmn}_{(\rm trans)}  = \frac{1}{2\mu_{(\bot)}}\left\{g^{ikmn} + \left[\varepsilon_{(\bot)} \mu_{(\bot)} {-}1 \right]
\left[g^{im} U^k U^n {-} g^{in} U^k U^m {+} g^{kn} U^i U^m  {-} g^{km} U^i U^n  \right]\right\} \,.
\end{equation}
In fact, there are now only four effective coupling constants $\mu_{(||)}$, $\mu_{(\bot)}$, $\varepsilon_{(||)}$, $\varepsilon_{(\bot)}$, defined as
$$
\frac{1}{2\mu_{(||)}} \equiv \Omega_1 +\Omega_2 \,, \quad
\frac12 \varepsilon_{(||)}  \equiv \Omega_1 +\Omega_2 +\Omega_3+\Omega_4+\Omega_5+\Omega_6 +\Omega_7+ 4 \Omega_8 \,,
$$
\begin{equation}\label{mnc}
\frac{1}{2\mu_{(\bot)}} \equiv \Omega_1 \,, \quad
\frac12 \varepsilon_{(\bot)}  \equiv \Omega_1 +\Omega_4+\Omega_6 \,,
\end{equation}
which play the roles of color permittivities, longitudinal and transversal, respectively.

\subsection{How does the model of unit dynamic vector field
\\ appear from the model of color aether?}

When the multiplet of vector fields $U^i_{(a)}$ converts into the set of parallel fields $U^i_{(a)} = q_{(a)} U^i$, so that $U^i$ becomes the unit dynamic vector field associated with the velocity four-vector of the aether, one can say, that the model of the SU(N)-symmetric aether transforms into the {\it extended} version of the Einstein-Yang-Mills-aether model (not into the simple version described above). In this case we deal with reduced master equations, and our first purpose is to write the master equation for the unit dynamic vector field $U^i$.
In order to obtain this equation we calculate the convolution of (\ref{CU1}) with $q^{(a)}$, and use the decompositions (\ref{P3})-(\ref{P5}), (\ref{A12})-(\ref{A14}) and (\ref{9})- (\ref{mn}).
The corresponding equation can be written in the form
$$
\nabla_i \left[{\cal K}^{imjn}_{(\rm long)} \nabla_m U_n \right] = \lambda U^j - \frac12 \D_i \left[{\cal A}_{(\rm long)}^{[mn]ij} F^{(b)}_{mn}q_{(b)} \right] +
{\cal K}^{imjn}_{(\rm trans)} U_n (\D_m q_{(a)})(\D_i q^{(a)}) +
$$
$$
+ \frac12 q^{(a)} {\cal K}^{ikmnj}_{(c)(b)(a)}   \left[q^{(c)}\nabla_i U_m + U_m \D_i q^{(c)}\right] \left[q^{(b)}\nabla_k U_n + U_n \D_k q^{(b)}\right] +
$$
\begin{equation}\label{0Red1}
+ \frac{\kappa}{2}q^{(a)}{\cal A}^{[ik]mnj}_{(c)(b)(a)} F^{(c)}_{ik} \left[q^{(b)}\nabla_m U_n + U_n \D_m q^{(b)}\right]  +
\frac{\kappa}{4}q^{(a)} {\cal C}^{ikmnj}_{(c)(b)(a)} F^{(c)}_{ik} F^{(b)}_{mn}
\,.
\end{equation}
Left-hand side of this equation contains the second order covariant derivative of the aether velocity four-vector $U_m$; in the right-hand side there are color vector $q_{(a)}$ and its gauge covariant derivative of the first order, $\D_m q^{(b)}$; the tensors ${\cal K}^{ikmnj}_{(c)(b)(a)}$, ${\cal A}^{[ik]mnj}_{(c)(b)(a)}$ and
${\cal C}^{ikmnj}_{(c)(b)(a)}$ are defined in (\ref{CU31}), (\ref{CU32}) and (\ref{CU33}), respectively.

The master equations for the color vectors $q^{(a)}$ can be obtained by convolution of (\ref{CU1}) with the projector $\Pi^{(a)(h)}$; they have the form
$$
\D_i \left[{\cal K}^{imjn}_{(\rm trans)} U_n \D_m q^{(h)}\right] = - \frac12 \D_i \left[{\cal A}_{(\rm trans)}^{[mn]ij} \Pi^{(h)}_{(b)}F^{(b)}_{mn} \right]-
$$
$$
{-} q^{(h)}\left[{\cal K}^{imjn}_{(\rm trans)}U_n \D_m q_{(a)}\D_i q^{(a)} {+} \frac12 {\cal A}_{(\rm trans)}^{[mn]ij} F^{(b)}_{mn}\D_i q_{(b)}  \right]
{-} \D_i q^{(h)} \left[{\cal K}^{imjn}_{(\rm long)} \nabla_m U_n {+} \frac12 {\cal A}_{(\rm long)}^{[mn]ij} F^{(b)}_{mn}q_{(b)} \right]
$$
$$
{+}\Pi^{(a)(h)} \left\{\frac12 {\cal K}^{ikmnj}_{(c)(b)(a)}   \left[q^{(c)}\nabla_i U_m {+} U_m \D_i q^{(c)}\right] \left[q^{(b)}\nabla_k U_n {+} U_n \D_k q^{(b)}\right] {+} \right.
$$
\begin{equation}\label{0Red11}
\left. {+}\frac{\kappa}{2}{\cal A}^{[ik]mnj}_{(c)(b)(a)} F^{(c)}_{ik} \left[q^{(b)}\nabla_m U_n {+} U_n \D_m q^{(b)}\right] {+}
\frac{\kappa}{4} {\cal C}^{ikmnj}_{(c)(b)(a)} F^{(c)}_{ik} F^{(b)}_{mn}
 \right\} \,.
\end{equation}
The left-hand side of this equation contains the second order gauge covariant derivative of the color vector $q_{(a)}$. Thus, (\ref{0Red1}) and (\ref{0Red11}) give the set of coupled equations of the second order in derivatives for the evolution of the unit vector $U^i$ and color vector $q^{(a)}$.

The equations for the gauge field have the form (\ref{Col1}), but now we obtain the reduced quantities:
$$
H^{ik}_{(a)} = q_{(a)} {\cal A}^{[ik]mn}_{(\rm long)} \nabla_m U_n + {\cal A}^{[ik]mn}_{(\rm trans)} U_n \D_m q_{(a)}  +
$$
\begin{equation}\label{Red1}
+ {\cal C}^{ikmn}_{(\rm trans)} F_{(a)mn} + q_{(a)} F^{(b)}_{mn} \ q_{(b)} \left[{\cal C}^{ikmn}_{(\rm long)}-{\cal C}^{ikmn}_{(\rm trans)} \right] \,,
\end{equation}
\begin{equation}\label{Red2}
\Gamma^i_{(a)} = {\cal G} f_{(a)(b)(c)} q^{(c)} U_k \left[\frac{1}{\kappa}  {\cal K}^{imkn}_{(\rm trans)} U_n \D_m q^{(b)} +
\frac12 {\cal A}^{[mn]ik}_{(\rm trans)} F^{(b)}_{mn}  \right]
 \,.
\end{equation}
Similarly, one can obtain reduced equations for the gravity field.

\subsection{The model with gauge covariant constant color vectors $q^{(a)}$}

The equations (\ref{0Red1}) can be simplified significantly, when $D_m q^{(a)}=0$, i.e., when the color vector $q^{(a)}$ is gauge-covariant constant normalized by unity, $q_{(a)}q^{(a)}=1$.
Clearly, it is possible, when the corresponding reduced equation (\ref{0Red11}) is satisfied:
$$
\D_i \left[{\cal A}_{(\rm trans)}^{[mn]ij} \Pi^{(h)}_{(b)}F^{(b)}_{mn} \right] =
$$
\begin{equation}\label{00Red11}
= \Pi^{(a)(h)} \left\{{\cal K}^{ikmnj}_{(c)(b)(a)}  q^{(c)} q^{(b)} \nabla_i U_m  \nabla_k U_n  +  \kappa {\cal A}^{[ik]mnj}_{(c)(b)(a)} q^{(b)} F^{(c)}_{ik} \nabla_m U_n   +
\frac{\kappa}{2} {\cal C}^{ikmnj}_{(c)(b)(a)} F^{(c)}_{ik} F^{(b)}_{mn}
 \right\} \,.
\end{equation}
In particular, this equation can be satisfied for arbitrary $U^i$ and $F_{mn}^{(a)}$, when the following equalities take place:
$$
{\cal A}_{(\rm trans)}^{[mn]ij} = 0 \,, \quad  \Pi^{(a)(h)} {\cal K}^{ikmnj}_{(c)(b)(a)}  q^{(c)} q^{(b)} =0 \,,
$$
\begin{equation}\label{000Red11}
\Pi^{(a)(h)} {\cal A}^{[ik]mnj}_{(c)(b)(a)} q^{(b)} = 0 \,, \quad  \Pi^{(a)(h)} {\cal C}^{ikmnj}_{(c)(b)(a)} = 0 \,.
\end{equation}
In fact, these requirements restrict the phenomenological parameters only; for instance, ${\cal A}_{(\rm trans)}^{[mn]ij} = 0$, when $\omega_1=0$ and $\omega_8+\omega_{14}=0$ (see (\ref{A14})).
Then the equation for the aether velocity four-vector takes the form
$$
\nabla_i \left[{\cal K}^{imjn}_{(\rm long)} \nabla_m U_n \right] = \lambda U^j - \frac12 \D_i \left[{\cal A}_{(\rm long)}^{[mn]ij} F^{(b)}_{mn}q_{(b)} \right] +
\frac12 q^{(a)}q^{(c)} q^{(b)} {\cal K}^{ikmnj}_{(c)(b)(a)}   \nabla_i U_m  \nabla_k U_n  +
$$
\begin{equation}\label{01Red1}
+ \frac{\kappa}{2}q^{(a)}q^{(b)}{\cal A}^{[ik]mnj}_{(c)(b)(a)} F^{(c)}_{ik} \nabla_m U_n   +
\frac{\kappa}{4}q^{(a)} {\cal C}^{ikmnj}_{(c)(b)(a)} F^{(c)}_{ik} F^{(b)}_{mn}
\,.
\end{equation}
However, the most serious information can be obtained from the integrability conditions for the equation $D_m q^{(a)}=0$. Indeed, if we start with the equation
\begin{equation}
\partial_m q^{(a)} = - {\cal G} f^{(a)}_{\ \ (b)(c)} A^{(b)}_m q^{(c)}  \,,
\label{v1}
\end{equation}
use the identity
\begin{equation}
\partial_{[n} \partial_{m]} q^{(a)} = 0 \,,
\label{v2}
\end{equation}
and the Jacobi identity (\ref{jfabc}), we obtain directly the first integrability condition
\begin{equation}
f^{(a)}_{\ \ (b)(c)} q^{(c)} F^{(b)}_{mn} = 0 \,.
\label{v3}
\end{equation}
As a consequence, we have to require that only the longitudinal component of $F^{(h)}_{mn}$ is non-vanishing:
\begin{equation}\label{99R}
\Pi^{(b)}_{(h)} F^{(h)}_{mn} = 0  \ \ \leftarrow \rightarrow \ \ F^{(b)}_{mn} = q^{(b)} q_{(h)} F^{(h)}_{mn}  \,.
\end{equation}
There is a trivial example of this symmetry: this condition is satisfied, when $F^{(b)}_{mn} = q^{(b)}F_{mn}$; then using (\ref{v1}) we see that
\begin{equation}
A^{(b)}_m = q^{(b)} A_m \,, \quad \partial_m q^{(a)} = 0 \,.
\label{v4}
\end{equation}
In other words, in that case we are faced with the quasi-Abelian  model with parallel potentials of the gauge field, and with the gauge-covariant constant vector $q^{(a)}$. For sure, this example is not unique, and in the nearest future we hope to consider in detail some cosmological applications of non-Abelian models of the spontaneously polarized color aether.

\section{Conclusions}

\noindent
1.
We established the theory of SU(N)-symmetric dynamic aether, i.e., based on the variation formalism we obtained the coupled system of master equations for the gauge, gravitational fields, and for the multiplet of vector fields, as well as, we presented the full-format catalog of constitutive tensors, appeared in the theory in the framework of the second order version of effective field theory.

\vspace{3mm}
\noindent
2.
We have shown that the standard dynamic aether, which is characterized by the single unit vector field, can appear from the established theory in the assumption that there exist a mechanism of spontaneous color polarization. This mechanism provides the color vector fields from the SU(N)-symmetric multiplet to become parallel in the group space, thus organizing a specific global direction in the four-dimensional space-time, described by the unit time-like four-vector, the aether velocity. We expect that the group (color) space, attributed to the SU(N)-symmetric aether model, is anisotropic (uni-axial or bi-axial), and the hypothetical phase transition in the color dynamic aether is accompanied by spontaneous color poling analogous to phenomena in electric and magnetic materials.

\vspace{3mm}
\noindent
3. In this paper we formulated the formalism, master equations of the new theoretical model and the ansatz about spontaneous color polarization; in the nearest future we hope to consider cosmological applications of this model, and to clarify the physical sense of the mechanism of a spontaneous color polarization in the SU(N)-symmetric dynamic aether.

\vspace{5mm}
\noindent
{\bf Acknowledgments}

\noindent
The work was supported by the Program of Competitive Growth of Kazan Federal University.

\end{document}